\documentstyle[12pt,epsfig]{article}

\advance\textheight by \topskip
\newcommand{\be}{\begin{equation}}
\newcommand{\ee}{\end{equation}}
\newcommand{\br}{\begin{eqnarray}}
\newcommand{\bea}{\begin{eqnarray}}
\newcommand{\beanon}{\begin{eqnarray*}}
\newcommand{\er}{\end{eqnarray}}
\newcommand{\eea}{\end{eqnarray}}
\newcommand{\eeanon}{\end{eqnarray*}}
\newcommand{\ba}{\begin{array}}
\newcommand{\ea}{\end{array}}
\newcommand{\bi}{\begin{itemize}}
\newcommand{\ei}{\end{itemize}}
\newcommand{\bn}{\begin{enumerate}}
\newcommand{\en}{\end{enumerate}}
\newcommand{\bc}{\begin{center}}
\newcommand{\ec}{\end{center}}

\newcommand{\ar}{\rightarrow}

\newcommand{\Dir}{\kern -6.4pt\Big{/}}%su lettere italiane minuscole
\newcommand{\Dirin}{\kern -10.4pt\Big{/}\kern 4.4pt}
\newcommand{\DDir}{\kern -7.6pt\Big{/}}%su lettere italiane maiuscole
\newcommand{\DGir}{\kern -6.0pt\Big{/}}%su lettere greche
\def\Ord{\buildrel{\scriptscriptstyle <}\over{\scriptscriptstyle\sim}}
\def\sm{\ifmmode{{\cal {SM}}}\else{${\cal {SM}}$}\fi}
\def\mt{\ifmmode{{m_{t}}}\else{${m_{t}}$}\fi}
\def\MH{\ifmmode{{M_{H}}}\else{${M_{H}}$}\fi}
\def\MWpm{\ifmmode{{M_{W^\pm}}}\else{${M_{W^\pm}}$}\fi}
\def\Wpm{\ifmmode{{{W^\pm}}}\else{${{W^\pm}}$}\fi}
\def\pl #1 #2 #3 {{\it Phys.~Lett.} {\bf#1} (#2) #3}
\def\np #1 #2 #3 {{\it Nucl.~Phys.} {\bf#1} (#2) #3}
\def\zp #1 #2 #3 {{\it Z.~Phys.} {\bf#1} (#2) #3}
\def\pr #1 #2 #3 {{\it Phys.~Rev.} {\bf#1} (#2) #3}
\def\prep #1 #2 #3 {{\it Phys.~Rep.} {\bf#1} (#2) #3}
\def\prl #1 #2 #3 {{\it Phys.~Rev.~Lett.} {\bf#1} (#2) #3}
\def\mpl #1 #2 #3 {{\it Mod.~Phys.~Lett.} {\bf#1} (#2) #3}
\def\rmp #1 #2 #3 {{\it Rev. Mod. Phys.} {\bf#1} (#2) #3}
\def\xx #1 #2 #3 {{\bf#1}, (#2) #3}

\begin{document}
%\baselineskip 1.1truecm
\tolerance=100000
\thispagestyle{empty}
\setcounter{page}{0}

\begin{flushright}
{\large Cavendish--HEP--97/05}\\
{May 1997\hspace*{.5 truecm}}\\ 
{Revised July 1997\hspace*{.5 truecm}}\\ 
\end{flushright}

\vspace*{\fill}

\begin{center}
{\large \bf 
Single-top production in the $tW^\pm$ channel \\[0.35cm]
and Higgs signals via $H\ar W^+W^-$ \\[0.35cm]
at the Large Hadron Collider}\\[2.cm]
{\large Stefano 
Moretti\footnote{E-mail: 
moretti@hep.phy.cam.ac.uk}}\\[.3 cm]
{\it Cavendish Laboratory,
University of Cambridge,}\\
{\it Madingley Road,
Cambridge, CB3 0HE, United Kingdom.}\\[1cm]
\end{center}

\vspace*{\fill}

\begin{abstract}
{\normalsize
\noindent
In a recent study by Dittmar and Dreiner it was shown that with appropriate
selection cuts the
signature of events containing two charged leptons and missing energy
represents the best chance of detecting the Standard Model Higgs scalar
in the mass range between 155 and 180 GeV, the primary decay of the Higgs
being into pairs of charged gauge bosons. The largest background 
to this channel is due to irreducible $W^+W^- X$ production.
In the present paper we calculate the contribution of events of the type
$bg\ar tW^\pm\ar b W^+W^-\ar b \ell^{+}\ell^{'-}\nu_\ell\bar\nu_{\ell'}$,
which have not been considered yet within the new selection strategy.
We show that the yield of this background is rather large, at the level
of that produced by $W^+W^-$, $t\bar t$ or $tb W^\pm $ events and thus needs
to be incorporated in future experimental analyses. However, we find that
its inclusion will not spoil the possibilities of Higgs detection in the 
above mentioned channel at the Large Hadron Collider.}
\end{abstract}

\vspace*{\fill}
\newpage

\section*{1. Introduction and motivations} 

In a recent paper by Dittmar and Dreiner \cite{herbi} (see also
Ref.~\cite{herbitalk}) it was pointed out that 
the signature of events with two charged leptons and missing 
energy/momentum at the Large Hadron Collider (LHC) represents the best
chance of detecting the Standard Model (SM) Higgs boson in the mass range
155 GeV $\Ord M_H\Ord$ 180 GeV. In Ref.~\cite{herbi}, simple selection criteria
were outlined, which should allow one to extract the Higgs decay
channel $H\ar W^+W^-$ from the non-resonant $W^+W^- X$ production (where
$X$ represents possible additional particles in the final state) with a 
signal-to-background ratio of about one-to-one, thus allowing a 
$5\div10\sigma$ detection with only 5 inverse picobarns
of integrated luminosity ${\cal L}=\int L{dt}$. The appealing prospect is 
that this significance can be achieved in less than one year of running of the 
CERN machine at the initial low luminosity 
$L=10^{33}~\mbox{cm}^{-2}~\mbox{sec}^{-1}$. Indeed, this is a clear improvement
compared to the Higgs search strategy based on the decay mode $H\ar ZZ^*\ar$ 
four charged leptons, which was the detection channel exploited even in the 
most recent experimental simulations \cite{CMS,ATLAS} for the mentioned Higgs 
mass range. This is evident if one considers that in order to
disentangle a $5\sigma$ signal in the latter case at least 100 fb$^{-1}$
are required.

%The procedure advocated in Ref.~\cite{herbi} is based on the following
%reasoning.
%The $H\ar ZZ^*\ar \ell^{+}\ell^{-} \ell^{'+}\ell^{'-}$ channel 
%(where $\ell,\ell'=e,\mu$) is certainly 
%rather clean, as the isolated and high $p_T$ leptons produced in the 
%$Z$-decays are easily extractable from the huge hadronic activity.
%This allows one to reconstruct a narrow mass peak in the invariant mass
%of the four lepton system. 
%However, the total branching ratio of the SM Higgs boson
%via $ZZ^*$-pairs into electron and/or muon pairs is tiny, for two 
%reasons. On the one hand,
%for Higgs masses in the range between 155 and 180 GeV the (off-shell) 
%$ZZ^*$ branching ratio is between $8\%$ and $6\%$, respectively. 
%On the other hand, $Z$-bosons decay into electrons/muons only three times out 
%of one hundred. In contrast, the $H\ar W^+W^-$ decay rate in the above mass
%window varies correspondingly between $78\%$ and $93\%$, with a peak
%around $M_H=170$ GeV, where BR$(H\ar W^+W^-)\approx97\%$. Furthermore,
%the leptonic channels $W^\pm\ar \ell\nu_\ell$ have a decay fraction of 
%$\approx11\%$ for each lepton flavour $\ell$.
%Therefore, the Higgs into two leptons plus missing energy decay 
%rate can be more than 200 times larger than that into four charged leptons,
%and yet two lepton tags can be performed. In the end,
%it is not surprising that such a big factor can 
%compensate for the absence of a narrow mass peak, because of the two neutrinos 
%escaping the detectors.

The first studies of the $H\ar W^+W^-$ decay mode 
\cite{ref9,ref10} in the context of Higgs searches at the LHC date back
to Ref.~\cite{ref13} and to the 1990 Workshop \cite{LHC} 
for a LHC  with $\sqrt s=16$ TeV.
Further analyses were subsequently performed, in Ref.~\cite{ref14}.
In various instances, also several signal-to-background studies were carried 
out (see Section 2 of Ref.~\cite{herbi} for a review). The unanimous conclusion
was that the $H\ar W^{+}W^{-} \ar
\ell^{+}\ell^{'-}\nu_\ell\bar\nu_{\ell'}$ channel (with $\ell,\ell'=e,\mu$) 
should provide a
useful tool to detect the Higgs boson in the mentioned mass range, though
more appropriate analyses (including hadronisation and detector
effects) were recognized to be needed to support those (mostly parton level)
results. 

This was done in Ref.~\cite{herbi}, by using the Monte Carlo (MC) program
PYTHIA \cite{ref15}. Further refinements were also introduced there,
which were not included in the previous literature. Namely, (i) the
inclusion of $W^\pm\ar \tau^\pm \nu_\tau\ar \ell^\pm\nu_\ell\nu_\tau$
decays (with $\ell=e,\mu$); 
(ii) the simulation of the background due to $gg\ar tb W^\pm$
events \cite{aa,tbW}; (iii) cuts previously employed \cite{ref13,ref14} 
were further supported by new constraints, introduced mainly in order 
to discriminate against the `irreducible' background from continuum 
production of $W^+W^- X$ events.

It is the purpose of this letter to provide additional material to 
motivate the exploitation of the $H\ar W^+W^-$ channel in Higgs searches at
the LHC, as we have studied the irreducible background 
due to
\be\label{bg_tW}
bg\ar tW^\pm\ar bW^+W^-\ar b \ell^{+}\ell^{'-}\nu_\ell{\bar\nu_{\ell'}}
\oplus\mbox{C.C.},
\ee
`single-top' events via $bg$-fusion (also called `$tW^\pm$-production'), 
which was not considered in Ref.~\cite{herbi}, and we will show that
this can be reduced to a manageable level by the same selection criteria
recommended in \cite{herbi}. 
In fact, for completeness, we have also computed the yield of the process
\be\label{bg_fusion}
bg\ar bW^+W^-\ar b \ell^{+}\ell^{'-}\nu_\ell{\bar\nu_{\ell'}}
\oplus\mbox{C.C.},
\ee
involving all the tree-level graphs producing the final state 
$b \ell^{+}\ell^{'-}\nu_\ell{\bar\nu_{\ell'}}$: that is, not only
the single-top ones isolated in reaction 
(\ref{bg_tW}) but also all the other diagrams contributing at the 
perturbative order ${\cal O}(\alpha_{em}^4\alpha_s)$\footnote{The symbol 
$\oplus~\mbox{C.C.}$ means that 
we have calculated also the charged conjugated processes initiated
by $\bar bg$-scatterings and these are included in 
all event rates presented in the following Sections.}.

With respect to the analysis performed in Ref.~\cite{herbi}, we will adopt
two simplifications, which we believe will not spoil the validity of our
conclusions. First, although 
we will implement the same cuts considered in Ref.~\cite{herbi},
we will confine ourself to the parton level only. However, since at
lowest order the final states of reactions (\ref{bg_tW})--(\ref{bg_fusion}) 
involve only one hadronic system (i.e., the $b$-quark fragmenting into hadrons)
whereas the Higgs signal $H\ar W^+W^-\ar \ell^{+}\ell^{'-}
\nu_\ell{\bar\nu_{\ell'}}$ is purely leptonic, we expect the effects of
hadronisation not to modify drastically the parton level dynamics. 
Second, we will only discuss the channels
$W^+W^- \ar \ell^{+}\ell^{'-}\nu_\ell{\bar\nu_{\ell'}}$ with 
$\ell,\ell'=e,\mu$, thus neglecting the case of $W^\pm$-decays into
tau leptons via the three-body channels 
$W^\pm \ar \tau^\pm \nu_\tau\ar \ell^\pm\nu_\ell\nu_\tau$. This is done
to simplify the description at parton level (especially in the case
of the complete process (\ref{bg_fusion})), as in this way we can avoid to 
calculate complicated two-to-seven and two-to-nine body subprocesses. In 
practice, contributions involving $\tau$-decays amount to $\approx1.9\%$
of the total $\approx7\%$ leptonic branching ratio of $W^+W^-$-pairs, so that
the bulk of the produced $W^+W^-$ events are indeed included in our study.
In general, we stress that we are here
only interested in the relative rates of signal and background and we expect
that the implementation of a full Monte Carlo simulation and the
inclusion of the $W^\pm\ar\tau\nu_\tau$
decays will presumably affect both in a rather similar manner. 

The reason for studying processes (\ref{bg_tW})--(\ref{bg_fusion}) as a
potential background in Higgs searches in the two leptons plus missing energy
channel is that single-top production via process
(\ref{bg_tW}) has very large event rates at the LHC, as its 
total cross section amounts to $55-60$ pb at $\sqrt s=14$ TeV (see later on), 
thus being comparable to that of the process $gg\ar tbW^\pm$ considered 
in Ref.~\cite{herbi} (see, e.g., Ref.~\cite{tbW,Heinson})\footnote{Note that
the leading order (LO) rates of the $gg\ar H$ signal for 155 GeV 
$\Ord M_H\Ord$ 180 GeV vary between 10 and 8 picobarns, approximately.}.
Furthermore, we stress that compared to the 
final state $tbW^\pm$, which eventually yields the signature $b\bar b W^+W^-$, 
that of reaction (\ref{bg_tW}) (and, more generally, of the complete
process (\ref{bg_fusion})) can boast only one additional particle with respect
to the Higgs signature (this rendering its reduction less
effective than that of $tbW^\pm$ events, which have two additional
jets\footnote{In this respect, we should mention that an extensive
compilation and a detailed discussion of processes involving single-top 
production at hadron colliders has recently been given \cite{tbW}.
In particular, according to the classification of Ref.~\cite{tbW},
there are six of these different hard parton scatterings. However,
process (\ref{bg_tW}) is the only one contributing at lowest order
to the irreducible background $W^+W^- X$ with one
additional particle in the final state (i.e., $X\equiv b$), as the
others always produce a second (light) jet.}). In fact, the latter 
is produced at lowest order via gluon-gluon
fusion into an on-shell Higgs boson, through a top quark loop \cite{xggh}:
$gg\ar H\ar W^+W^-\ar \ell^{+}\ell^{'-}\nu_\ell{\bar\nu_{\ell'}}$. However,
we notice that the $K$-factor of Higgs production via
$gg$-fusion has been shown to be very large, around two 
\cite{Kfacgg1,Kfacgg2,Kfacgg3} in the mass range 
155 GeV $\Ord M_H\Ord$ 180 GeV (and outside, as well 
\cite{allspectrum}). 
In particular, a large component of the next-to-leading (NLO) order corrections
to the gluon-gluon fusion mechanism of Higgs production is due to the real 
radiation \cite{allspectrum} of a quark or gluon, so that also 
signal events are naturally accompanied by an additional detectable jet inside 
the detectors.

For reference, we recall that the matrix element of
process (\ref{bg_tW}) was already computed in Ref.~\cite{ME} and first
studied in the context of Higgs
searches (and of $W^+W^-$ physics, as well) in Ref.~\cite{singlet}
(for its relevance in the case of top-quark physics, see 
Ref.~\cite{tbW}). However, only the invariant mass region
$M_{tW}\equiv \sqrt{\mathaccent94{s}}>850$ GeV was considered there, as a 
background to signatures of heavy Higgs bosons decaying into longitudinal 
polarised $W^+W^-$-pairs \cite{WWpolar}.

The plan of this paper is as follows. In the next Section we
give some details of the calculation.
Section 3 is devoted to a discussion of the results.
Our conclusions are in Section 4.

\section*{2. Calculation} 

The tree-level Feynman `topologies' that one needs for computing
processes (\ref{bg_tW})--(\ref{bg_fusion}) are given in Fig.~1. 
Once all the internal propagator are correctly inserted one gets a total of
43 Feynman graphs (the single-top diagrams pertaining to reaction
(\ref{bg_tW}) can be obtained from the topologies 2 and 3). To calculate the
corresponding amplitude squared we have used 
MadGraph \cite{tim} and HELAS
\cite{HELAS}. The integrations over the appropriate phase spaces have been 
performed by using {\tt VEGAS} \cite{VEGAS}. 
The codes produced have been carefully checked for gauge and BRS
\cite{BRS} invariance. Furthermore, the total cross section for process
(\ref{bg_tW}), obtained by selecting the only two graphs with on-shell
top production out of those displayed in Fig.~1, has been compared
against the results given in Ref.~\cite{tbW} for the Tevatron and in
Ref.~\cite{singlet} for the Superconducting Super Collider (SSC), with
identical choice of parameters, cuts (where applied) and structure
functions, and perfect agreement has been found.
The signal rates have been computed by using the program already adopted
in Refs.~\cite{update,io}. However, contrary to the case of 
Ref.~\cite{update} where NLO rates were used to calculate the 
Higgs production cross sections via $gg$-fusion, and in line with 
Refs.~\cite{herbi,io}, we have used here the LO results.
This has been done for consistency, as one-loop calculations 
do not exist to date for processes (\ref{bg_tW}) and (\ref{bg_fusion}).
It is however important to point out that whereas  
the complete corrections
to the main Higgs production process via gluon--gluon fusion 
are large and positive \cite{allspectrum} those
to the single-top process (\ref{bg_tW}) are expected to be much 
smaller \cite{tbW}.

The $b$-quark in the initial state of 
reactions (\ref{bg_tW})--(\ref{bg_fusion}) 
has been treated as a constituent of the proton with the appropriate
momentum fraction distribution $f_{b/p}(x,Q^2)$, as given by the 
parton distribution functions (PDFs).
As default set of the latter we have used MRS(A) \cite{MRSA}. 
However, as the PDFs of the gluon inside the 
proton are not so well know at medium and small $x$ and since those 
of $b$-quarks suffer from potentially large (theoretical) 
uncertainties
%\footnote{In fact, the $b$-sea distributions
%are not measured by experiment, rather these are obtained from the gluon
%distributions splitting into $b\bar b$ pairs by using the 
%Dokshitzer-Gribov-Lipatov-Altarelli-Parisi evolution equations \cite{DGLAP}. 
%In general, the dynamics of such evolution is different from set to set.
%For example, the two kinds of PDFs 
%considered here (i.e., the Martin-Roberts-Stirling and CTEQ various packages,
%see later on) treat differently the $b$-threshold region in the gluon
%splitting. Whereas the CTEQ collaboration 
%evolves the $b$-sea distribution beginning from $\mu=m_b$, the second group 
%starts the evolution at $\mu=2m_b$ \cite{tbW}.
%Note that $b$-quarks are not even included in the GRV sets of parton 
%distributions \cite{GRV94}.} 
(see, e.g., Ref.~\cite{MRSCHM}), we have produced our results in the case 
of other 4 sets of recent NLO structure functions, which give
excellent fits to a wide range of deep inelastic scattering data
(including the latest measurements from the  HERA $ep$ collider)
and to data on other hard scattering processes. These are the packages
 MRS(A',\- G,\- R1,\- R2) \cite{MRSA,MRSG,mrs96fit}. 
%%%%%%%%%%%% 
%MRS(105,\- 110, 115, 120, 125,\- 130) and CTEQ(2M,\- 2MS,\- 2MF,\- 2ML,\- 3M)
%\cite{MRSA,MRSG,mrs96fit,MRSalphas,CTEQ3}. 
%In each case the appropriate value of $\Lambda^{(n_f)}_{\overline{{MS}}}$
%(in the modified Minimal-Subtraction scheme) has been used. In particular,
%in the case
%of the MRS(A) set we have adopted $\Lambda^{(4)}_{\overline{{MS}}}=230$~MeV.
%%%%%%%%%%%%
The QCD strong coupling $\alpha_s$ entering explicitly in the
production cross sections and implicitly in the parton distributions has
been evaluated using the {\tt CERNLIB} package
%%%%%%%%%
%at two-loop order with five active quark flavours 
%and $\Lambda^{(n_f\ne4)}_{\overline{{MS}}}$ calculated according to the 
%prescriptions in Ref.~\cite{MARCIANO} and 
%%%%%%%%%
at the scale $\mu=\sqrt{\mathaccent94{s}}$ (i.e., 
the CM energy at parton level). We will discuss
the $\mu$ dependence of the LHC rates in the following Section. 
Indeed, one should remember that also
the value of $\alpha_s$ associated with each parton set 
represents in principle a residual
source of error in the predictions of the different PDFs. However, 
the value 
adopted in each set is chosen to match the data during the
extraction, so that we do not expect $\alpha_s$ to be a significant source of 
uncertainty. 

In the numerical calculations 
we have adopted the following values for the electromagnetic coupling constant
and the weak mixing angle:
$\alpha_{em}= 1/128$ and  $\sin^2\theta_W=0.2320$.
For the gauge boson masses and widths we have taken
$M_{Z}=91.1888$ GeV, $\Gamma_{Z}=2.5$ GeV,
$M_{W^\pm}\equiv M_{Z}\cos\theta_W\approx80$ GeV and
$\Gamma_{W^\pm}=2.08$ GeV, while for the top mass we have used
$m_t=175$ GeV \cite{value}. All other fermions have been considered
massless, including the $b$-quark. In particular, the choice $m_b=0$
has been maintained also in the Yukawa couplings of the theory. In this
way, no diagram involving radiation of Higgs bosons off the $b$-lines
has been included in process (\ref{bg_fusion}). For simplicity, we have
set the CKM matrix element of the top-bottom coupling equal to one. In this
respect, we recall again Ref.~\cite{tbW}, where it was shown that off-diagonal
CKM matrix element subprocesses are negligible at the Tevatron. We do expect 
the  same to occur at LHC regimes.

Finally, as total CM energy of the colliding beams at the 
LHC we have adopted the value $\sqrt s=14$ TeV.

\section*{3. Results}

Our results are presented in Tab.~I and Figs.~2--6. Note that
for the time being we assume that
no $b$-tagging identification is exploited in
events of the type (\ref{bg_tW})--(\ref{bg_fusion}). The integrated
luminosity adopted throughout the paper will be $5~\mbox{fb}^{-1}$.

\subsection*{3.1 Selection cuts}

As event selection procedure we have adopted the same one exploited in 
Ref.~\cite{herbi}, to which we refer the reader for a detailed discussion
concerning the meaning of the various cuts. We only tabulate
these here, in order to introduce a notation that will be used in the 
remainder of this paper (note that the two leptons $\ell$ and $\ell'$
must be of opposite sign). Following the same numerical 
sequence as in \cite{herbi}, we ask (at parton level):
\begin{enumerate}
\item $p_T^{\ell,\ell'}>10$ GeV, $p_T^{\ell}~\mbox{or}~p_T^{\ell'}>20$ GeV, 
$\theta_{\ell,\ell'}>10^{{\mbox{\tiny o}}}$, for the transverse momentum and the
separation angle of the two leptons;
\item $|\eta^{\ell,\ell'}|<2$, for the pseudorapidity of the two leptons;
\item $E_b<5$ GeV if $\theta_{b\ell,b\ell'}<20^{\mbox{\tiny o}}$, for the
energy of the $b$-quark and the separation angles between the leptons
and the $b$-quark;
\item $M_{\ell\ell'}<80$ GeV, for the dilepton mass;
\item $p_T^{\mbox{\tiny{miss}}}>20$ GeV, for the missing transverse momentum
of the event;
\item $\phi<135^{\mbox{\tiny o}}$, for the angle between the two leptons in the 
plane
transverse to the beam direction;
\item if $|\eta^b|<2.4$ then $p_T^b<20$ GeV, for the transverse momentum
and the pseudorapidity of the $b$-quark;
\item $|\cos\theta|<0.8$, for the cosine of the dilepton system with respect
to the beam direction;
\item the enforcement $10^{\mbox{\tiny o}}<\phi<45^{\mbox{\tiny o}}$, 
for the same angle defined in 6.;
\item $M>140$ GeV, for the estimated invariant mass of the $W^+W^-$-system;
\item $0<\cos\xi<0.3$, for the angle between the lepton with the 
largest transverse momentum, boosted to the dilepton rest frame, and the
momentum vector of the dilepton system.
\end{enumerate}

A few comments are in order before proceeding further, concerning the
application of cuts 3. and 7.
%%%%%%%%%%%%%%%%%
%In particular, whereas the
%constraints 1., 2., 4.--6. and 8.--11. can in our opinion 
%be applied in a straightforward
%way at parton level (in fact, we do not expect that photon radiation
%from the final state can significantly change the pattern of the
%acceptance fractions),
%the other two, 3. and 7., would more appropriately need to be implemented 
%at hadronic level, after the emission of initial and final state QCD
%radiation and the fragmentation of all partons produced (of course available
%in MC programs used for experimental simulations). 
%%%%%%%%%%%%%%%%%%
As for cut 3.,
according to our parton level implementation, background events are 
rejected if the $b$-quark is energetic and is found near one of the leptons. 
On the one hand, we certainly expect the $b$-quark to be very fast.
%(particularly when produced in a top-decay). 
On the other hand, we do not
see a priori any reasons why the quark and the leptons should be created 
in collinear configurations. 
This is in fact confirmed by the spectra given in Fig.~2a.
%%%%%%%%%%%%%%%%%%
%The $b$-quark is very energetic since on average $E_b\approx70$ GeV and
%the $b$-momentum is well separated from those of the leptons as the two 
%distributions in angle fall very sharply when $\theta_{b\ell,b\ell'}\ar0$.
%%%%%%%%%%%%%%%%%%
However, things would look quite different at hadron level.
In fact, the jet produced by the bottom quark would have a finite size and 
the hadrons produced in the showering would carry only a fraction of the
original parton energy. Although we miss these two aspects, we stress 
that the two systematics errors we introduce with our treatment do work
in opposite directions, so to counterbalance each other. 
%%%%%%%%%%%%%%%%%%%%
%That is, on 
%the one hand, the parton energy falling in the defined cone regions 
%around the leptons is much higher than that of the single hadron, on the
%other hand, such energy is concentrated all along one direction rather far
%apart from those of the leptons themselves. In the end, presumably, the
%overall effect will be a small overestimate of the accepted background rates,
%as the $b$-quark direction (roughly, the direction of the jet axis)
%is often in the same hemisphere with that of (one of) the leptons.
%%%%%%%%%%%%%%%%%%%%%
Cut 7. will have no effect on our signal 
rates, as we are considering here neither $\tau$-decay modes nor initial
state QCD radiation, whereas for the background it will act directly on the 
$b$-parton. This corresponds to an
overestimate of the signal, while we believe that the accepted 
fraction of background events will be predicted accurately, as
the efficiency in reconstructing the $b$-momentum from
the hadrons should be rather high because of the clean environment (the two 
leptons) in which the $b$-quark fragments. Whichever is the interplay
between parton and hadron level, is anyway clear that it is cut 7. that 
will introduce a strong reduction factor on the background, as 
the $b$-jet will be easily 
detectable in pseudorapidity and will also have a large
transverse momentum\footnote{Note that Fig.~2b has been plotted after having
already implemented the constraints 1.--6., and so will be in all forthcoming
Figures.} (see Fig.~2b).

\subsection*{3.2 Theoretical error}

As first step of our analysis we have compared the production rates of
process (\ref{bg_tW}) and (\ref{bg_fusion}) and found that in Higgs 
searches (that is, for the selection cuts 1.--11.) the additional 
contributions from the non-top diagrams of Fig.~1 are negligible. Therefore,
in the following we will neglect them.

As one of the possible means of estimating the uncertainty of the theoretical
predictions on the gluon distribution (and hence the $b$-one) we have 
calculated the cross section for on-shell single-top
production via the two-to-two body process $bg\ar tW^\pm\oplus\mbox{C.C.}$
for the mentioned five sets of PDFs. 
The spread around the value obtained from MRS(A) 
(the set that we will adopt as a default in the following)
is between $-9\%$ (from MRS(G)) and $+3\%$ (from MRS(R2)). 
This will represent throughout the paper the
conservative estimate at present time of the uncertainty on the
$bg$-fusion cross section into single-top quarks due to the
parton distributions. Note that the above values roughly
compare to those identified (for the same sets)
in Ref.~\cite{update} for the case of $gg$-fusion into an on-shell Higgs,
so that this helps in this context in carrying out a consistent 
signal-to-background analysis.

Finally, the factorisation scale dependence (which quantifies our ignorance of 
higher order corrections) of the background rates via process
(\ref{bg_tW}) has been estimated by varying $\mu$ in the range 
$\sqrt{\hat{s}}/2 < \mu < 2\sqrt{\hat{s}}$ when calculating 
the total cross section.
We notice that, using MRS(A), differences with respect
to the rate at $\mu=\sqrt{\hat{s}}$ are less than $0.1\%$ 
at $\mu=\sqrt{\hat{s}}/2$ and $-3\%$ at $\mu=2\sqrt{\hat{s}}$. 
We have verified that similar effects also occur when
other PDFs are used. Such numbers are rather small and presumably comparable 
with the experimental uncertainties\footnote{Note that the scale dependence 
of processes producing the final state $tW^\pm X$ at the Tevatron 
has been studied in Ref.~\cite{tbW}, where
variations between
$-14\%$ and $+20\%$ were quoted, for $\mu$ spanning over
the range between $\mu=m_t/2$ and $2m_t$.}. 

\subsection*{3.3 Kinematics and event rates}

One should expect the impact of the cuts 8. and 9. to be similar
on both signal and background, as can be noticed from Figs.~3 and 4, 
respectively. In fact, the shapes of the corresponding distributions are
almost identical\footnote{Please notice the arrow in Fig.~4 to indicate 
the maximum value of the signal at $\cos\phi\approx1$.}. 

Not even the invariant mass $M$ of the reconstructed 
(from the lepton and the missing momenta) ${W^+W^-}$-system is helpful to
discriminate the signal from the background (see Fig.~5). 
In fact, the background
spectrum is almost entirely beyond the minimum value of 140 GeV implied
by cut number 10. The discrimination power of such constraint is thus very 
limited, if not self-defeating.

The only cut among those introduced in Ref.~\cite{herbi} to reduce the
irreducible $W^+W^- X$ background from $W^+W^-$, $t\bar t$ and $tb W^\pm $
events which is also effective against $bg\ar tW^\pm$ 
is cut 11., as can be appreciated from Fig.~6. In fact, the two charged 
leptons from the background have a rather large angular spread, so that 
the maximum of the background distribution is located around the value 0.6.

The accepted event rates, for both signal and background, for
a selection of six Higgs masses, are presented in Tab.~I ($\sqrt s=14$ TeV
and ${\cal L}=5~\mbox{fb}^{-1}$). When comparing the numbers in Tab.~I
one should bear in mind that the background rates there should be added to
those given in Tab.~2 of Ref.~\cite{herbi}. This should however be done
after treating all background sources in $W^+W^- X$ events on the same footing
(i.e., consistently at parton or, better, hadron level). 
This is beyond our intentions and capabilities, 
our aim here is to make the point that background events from
process (\ref{bg_tW}) are large compared to the signal, as they vary
between $11\%$  and $22\%$ of the Higgs rates, depending on the mass of the
scalar. Therefore, their effect in the signal-to-background significance
is of the same order as that of any of the three processes $pp\ar W^+W^-$,
$pp\ar t\bar t$ and $pp\ar tb W^\pm $ studied in Ref.~\cite{herbi}, 
especially considering the fact that our parton level analysis overestimate
the signal by a factor of two (compare the numbers in our Tab.~I to those
in Tabs.~1--2 of Ref.~\cite{herbi} in response to the application
of cut number 7.), while more accurately predicting the background rates.
Finally, one should notice the effectiveness of the selection strategy based
on the cuts 1.--11. against events of the type (\ref{bg_tW}), as the overall
reduction factor on this background is above 1000 !

\section*{4. Summary and conclusions}

In this paper, we have studied the yield of process 
$bg\ar tW^\pm\ar bW^+W^-\ar b \ell^{+}\ell^{'-}\nu_\ell{\bar\nu_{\ell'}}$
(where $\ell,\ell'=e,\mu$) as `irreducible' background to the 
$gg\ar H\ar W^+W^-\ar \ell^{+}\ell^{'-}\nu_\ell{\bar\nu_{\ell'}}$ signature
of the Standard Model Higgs boson, which has recently been claimed
as the most viable channel to detect such a scalar in the mass range
155 GeV $\Ord M_H\Ord$ 180 GeV at the Large Hadron Collider. Although
we have confined ourselves to the parton level only, we believe we have 
performed a consistent signal-to-background analysis, exploiting the
same event selection procedure advocated in literature. (In particular, the
shape of the parton level distributions used to disentangle the signal 
from the irreducible $W^+W^- X$ noise resembles very closely those previously
obtained 
at hadron level). This has enabled us to assess that non-resonant $W^+W^- X$ 
events due to single-top production
via $bg$-fusion are rather numerous, and comparable to the rates of any of
the reactions 
$gg,q\bar q\ar t\bar t$, $gg\ar tb W^\pm $ and $gg,q\bar q\ar W^+W^-$, which
have in fact been shown to represent the largest components of the total
background to the Higgs detection channel in two charged leptons and 
missing energy/momentum. In contrast, $bg\ar bW^+W^-\ar 
b\ell^{+}\ell^{'-}\nu_\ell{\bar\nu_{\ell'}}$ events not proceeding
via single-top diagrams are negligible. 
Therefore, we think that the production process $bg\ar tW^\pm$ that we have
studied should be included in the experimental Monte Carlo simulations
which will be used in order to confirm or disprove the existence  of the
Higgs scalar of the Standard Model in the above mass range at the CERN
proton-proton collider. We believe
this to be
particularly important, as the discussed signature does not allow one to
reconstruct the narrow Higgs resonance (because of the neutrinos escaping 
the detectors). In fact, the presence of the latter will be established by an
`event counting' operation over a rather broad region in mass, where
the $H\ar W^+W^-$ signal and the $bg\ar tW^\pm$ background have a very
similar shape. For the purpose of aiding future analyses, we 
make available  upon request the 
electronic version of the matrix element for 
$bg\ar bW^+W^-\ar b\ell^+\ell^{'-}\nu_\ell{\bar{\nu_{\ell'}}}$.

However, we would like to conclude this study by stressing that the inclusion
of the single-top background in $tW^\pm$ events will certainly 
not spoil the chances
of detecting the Standard Model Higgs in the advocated decay channel, and that 
the exploitation of the two charged leptons and missing energy signature remains
crucial in Higgs searches at hadron colliders.

\section*{5. Acknowledgements}

We thank Bryan Webber for reading the manuscript and for useful comments
and Herbi Dreiner for his assistance with the bibliography of process 
(\ref{bg_tW}).
We are grateful to the UK PPARC for financial support.

\goodbreak

\vfill
\newpage

\subsection*{Table Captions}
\begin{description}

%\item{[I]   } Total cross sections for on-shell single-top production
%via $bg$-fusion at the LHC for: {\bf a}) sixteen different
%sets of parton distribution functions; {\bf b}) ten different values of the 
%scale $\mu$ (adopting the MRS(A) set).

\item{[I]  } The expected signal and background number of events for
5 fb$^{-1}$ at the LHC after the
application of the selection criteria discussed in the text. Only the two-body
decays $W^\pm\ar \ell\nu_\ell$ with $\ell=e,\mu$ are considered in both
signal and background. The parton distribution functions used are MRS(A). 

\end{description}

\subsection*{Figure Captions}
\begin{description}

\item{[1]   } Feynman diagram topologies contributing at tree-level
to the process $bg\ar b \ell^- {\ell'}^+\bar\nu_\ell\nu_{\ell'}$,
where $\ell$ represents a lepton. 
Internal wavy lines represent a photon, a $Z$ or a $W^\pm$, 
whereas the internal solid ones refer to a lepton, a neutrino, a bottom-
or a top-quark, as appropriate. The total number of graph is 43 (excluding
Higgs couplings). The single-top diagrams are number 2 and 3. Charge
conjugated diagrams can be trivially obtained by reversing the fermion lines.

\item{[2]   } Differential distributions in ({\bf a}) 
energy of the final state $b$-jet
(left plot) and its angular separation from the two leptons (right plot:
from that generated by the top-decay $W^\pm$, solid line, and
from that generated by the non-top-decay $W^\pm$, dashed line)
and ({\bf b}) transverse momentum (left plot)
and pseudorapidity (right plot) of the final state $b$-jet in events of the type
$bg\ar tW^\pm\ar bW^+W^- \ar b(\ell^-\bar\nu_\ell)({\ell'}^+\nu_{\ell'})$, 
with $\ell,\ell'=e,\mu$, at the LHC, ({\bf a}) before the
acceptance cuts and ({\bf b}) after the acceptance cuts 1--6. 
The parton distribution functions used are MRS(A). 

\item{[3]   } Differential distributions in the polar angle of the dilepton
system with respect to the beam direction for the Higgs signal 
($M_H=170$ GeV, solid histogram) and the single-top background (dashed
histogram) at the LHC after the acceptance cuts 1--6.
The parton distribution
functions used are MRS(A). Note that the background rates
have been divided by two in order to facilitate the comparison between
the two curves.

\item{[4]   } Differential distributions in the azimuthal angle of the dilepton
system in the plane transverse to the beam direction for the Higgs signal 
($M_H=170$ GeV, solid histogram) and the single-top background (dashed
histogram) at the LHC after the acceptance cuts 1--6.
The parton distribution
functions used are MRS(A). Note that the background rates
have been divided by two in order to facilitate the comparison between
the two curves.

\item{[5]   } Differential distributions in the estimated invariant mass of the
$\ell^-{\ell'}^+\bar\nu_\ell\nu_{\ell'}$ system for the Higgs signal 
($M_H=170$ GeV, solid histogram) and the single-top background (dashed
histogram) at the LHC after the acceptance cuts 1--6.
The parton distribution
functions used are MRS(A). Note that the background rates
have been divided by two in order to facilitate the comparison between
the two curves.

\item{[6]   } Differential distributions in the angle
between the lepton with highest transverse momentum, boosted to the dilepton
rest frame, and the momentum vector of the dilepton
system itself for the Higgs signal 
($M_H=170$ GeV, solid histogram) and the single-top background (dashed
histogram) at the LHC after the acceptance cuts 1--6.
The parton distribution
functions used are MRS(A). Note that the background rates
have been divided by two in order to facilitate the comparison between
the two curves.

\end{description}

\vfill
\newpage

\begin{table}%[p]%[htbp]
\begin{center}
\begin{tabular}{|c|c|c|c|c|c|c|c|}
\hline
\multicolumn{8}{|c|}
{\rule[0cm]{0cm}{0cm}
Accepted event rates}
\\ \hline
\rule[0cm]{0cm}{0cm}
process $pp\ar X$ &\omit &\omit  &\omit   &\omit\quad $N_{\mbox{\tiny{ev}}}$ &
\omit  &\omit  & \\ \hline\hline
\rule[0cm]{0cm}{0cm}
                              & No cut & cut 1--3 & cut 4--6 & cut 7 & 
                                cut 8--9 & cut 10 & cut 11 \\ \hline
$gg\ar H\ar W^+W^-$           & & & & & & & \\ 
$M_H=155$ GeV                 & 1832 & 1032 &  893 &  893 & 209 & 147 &  70\\ 
$M_H=160$ GeV                 & 2002 & 1154 & 1035 & 1035 & 267 & 208 & 109\\ 
$M_H=165$ GeV                 & 2017 & 1179 & 1054 & 1054 & 270 & 219 & 119\\ 
$M_H=170$ GeV                 & 1929 & 1141 &  988 &  988 & 244 & 199 &  99\\ 
$M_H=175$ GeV                 & 1829 & 1087 &  901 &  901 & 215 & 176 &  79\\ 
$M_H=180$ GeV                 & 1702 & 1019 &  801 &  801 & 184 & 152 &  61\\ 
\hline
$bg\ar tW^\pm\ar bW^+W^-$     &13408 & 8254 & 2794 &  238 &  44 &  42 &  13\\ 
\hline\hline
\multicolumn{8}{|c|}
{\rule[0cm]{0cm}{0cm}
$\sqrt s=14$ TeV\qquad\qquad\qquad${\cal L}=5$ fb$^{-1}$\qquad\qquad\qquad 
$m_t=175$ GeV}
\\ \hline

%\multicolumn{8}{|c|}
%{\rule[0cm]{0cm}{0cm}
%MRS(A)}
%\\ \hline

%\multicolumn{8}{c}
%{\rule{0cm}{1.0cm}
%{\Large Tab. I}}  \\

\end{tabular}
\end{center}
\end{table}
\
\vskip-1.0cm
\centerline{\Large Tab. I}

\vfill
\clearpage

\begin{figure}[p]
~\epsfig{file=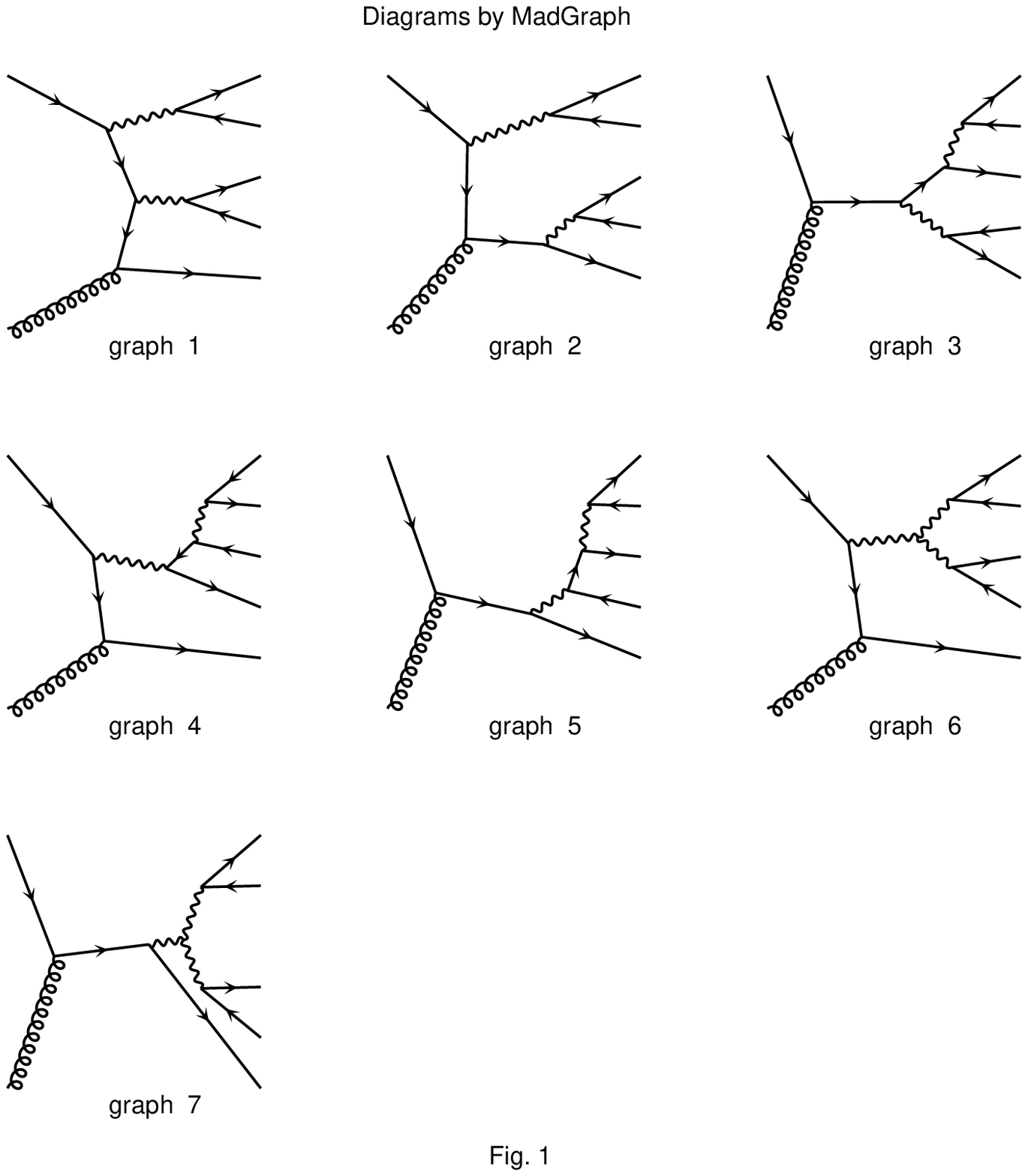,height=22cm}
\end{figure}
\stepcounter{figure}
\vfill
\clearpage

\begin{figure}[p]
~\epsfig{file=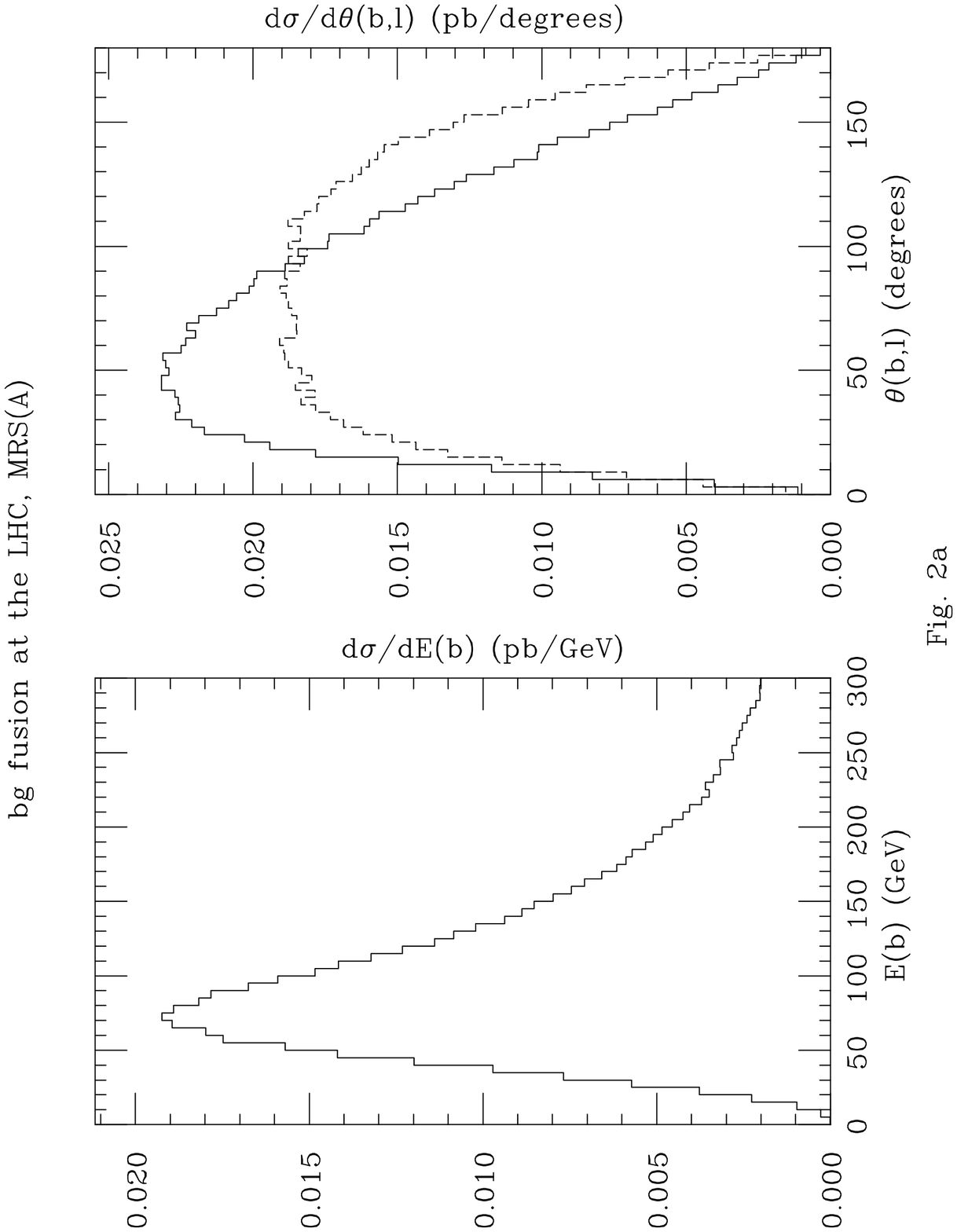,angle=180,height=22cm}
\end{figure}
\stepcounter{figure}
\vfill
\clearpage

\begin{figure}[p]
~\epsfig{file=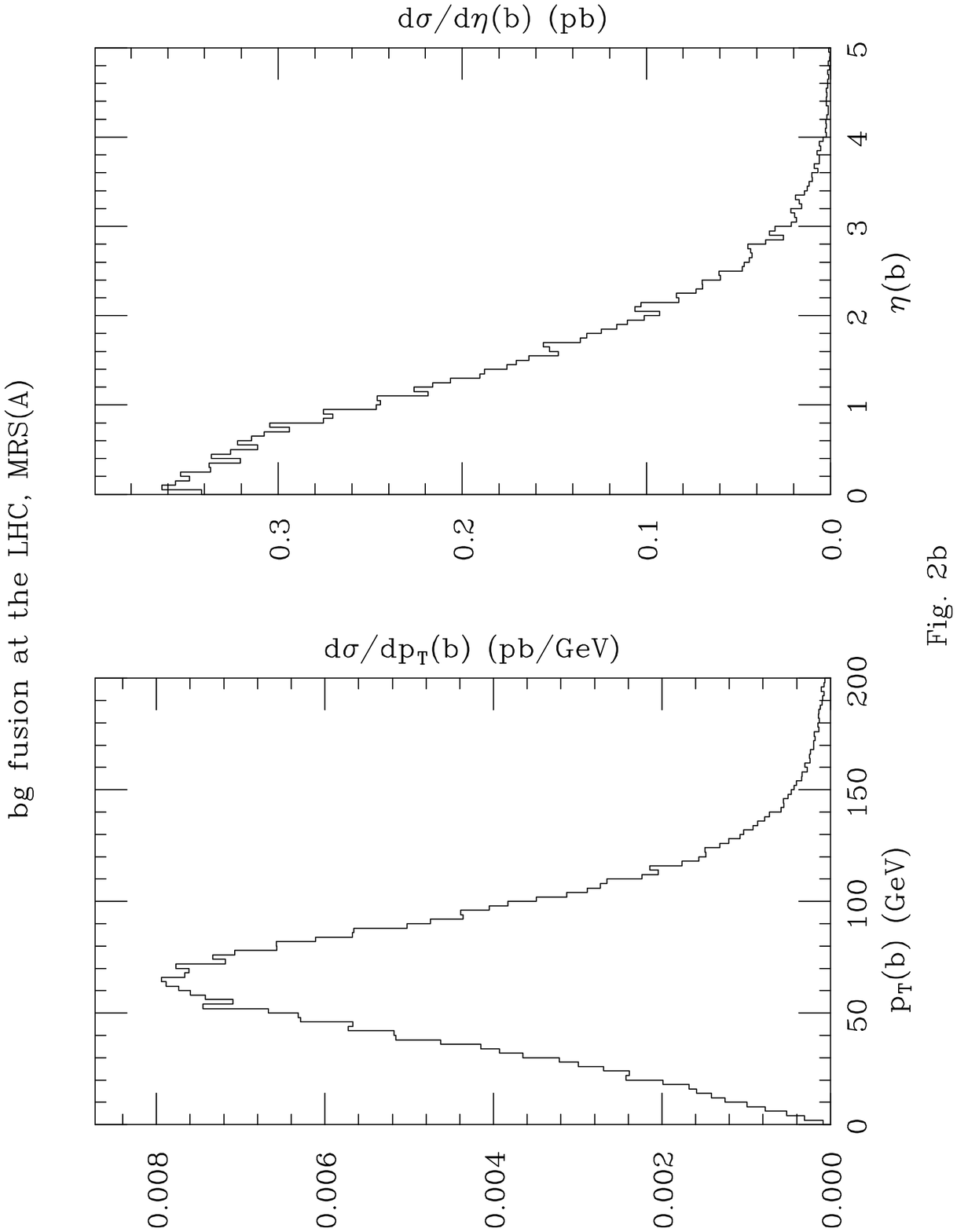,angle=180,height=22cm}
\end{figure}
\stepcounter{figure}
\vfill
\clearpage

\begin{figure}[p]
~\epsfig{file=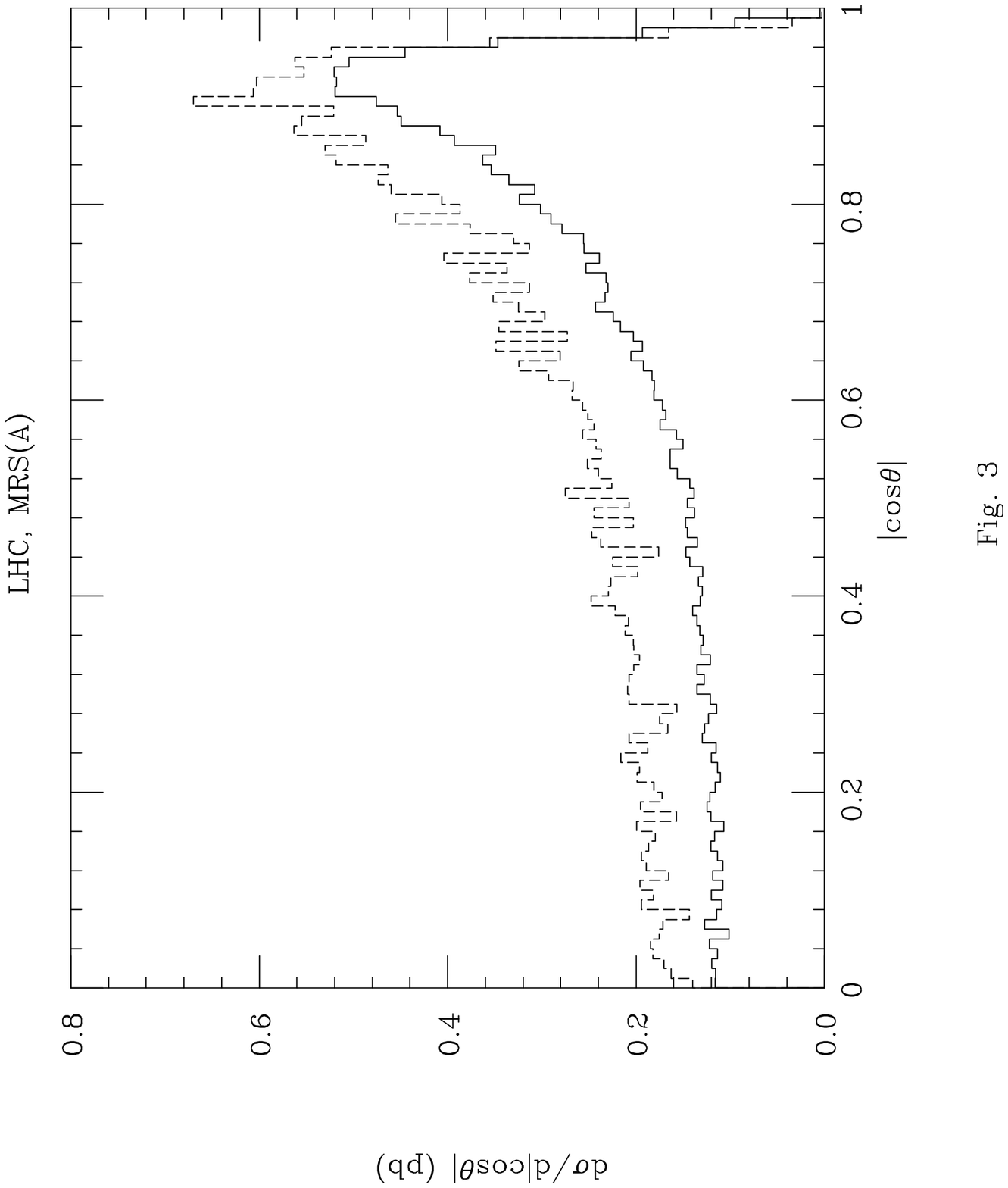,angle=180,height=22cm}
\end{figure}
\stepcounter{figure}
\vfill
\clearpage

\begin{figure}[p]
~\epsfig{file=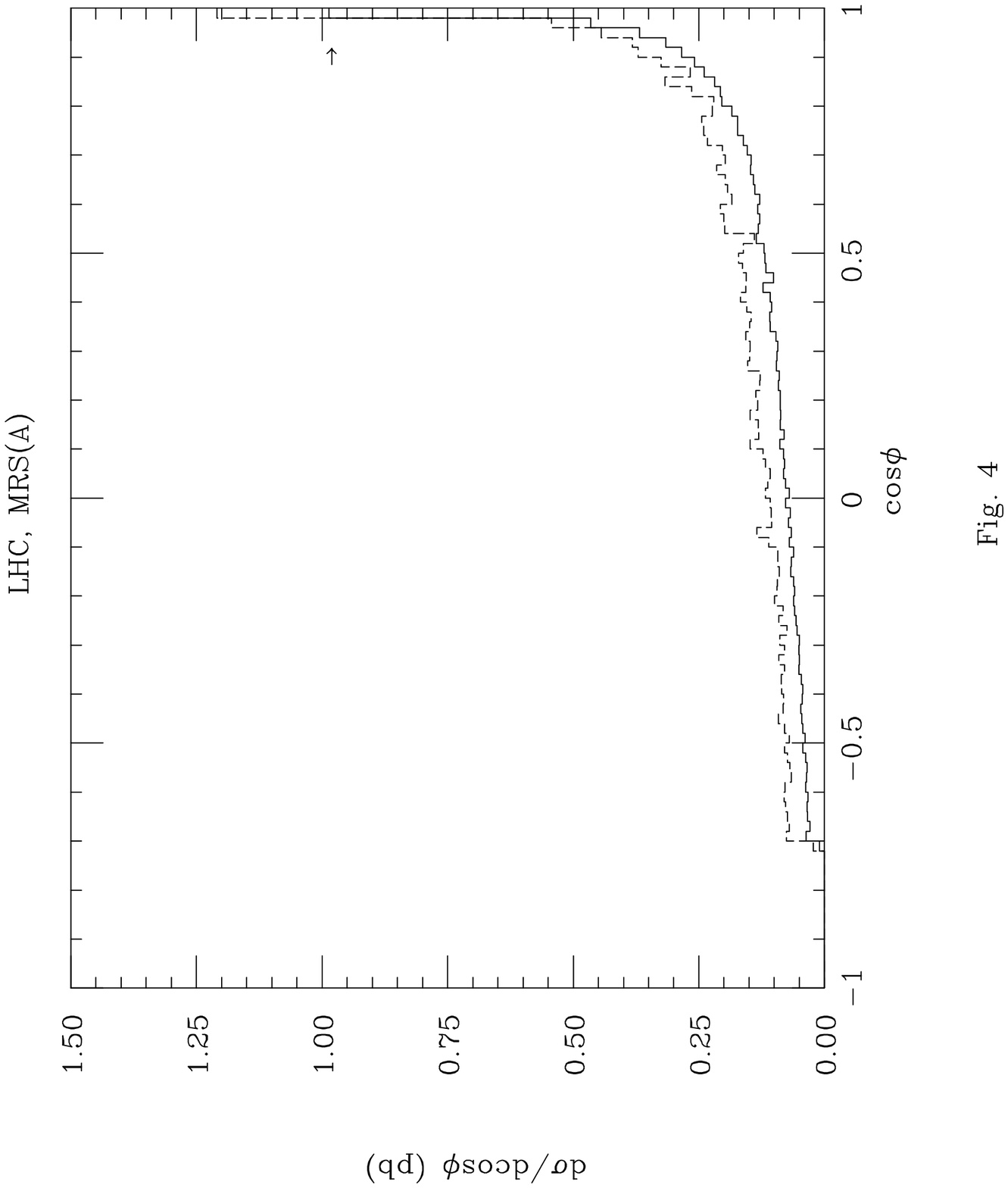,angle=180,height=22cm}
\end{figure}
\stepcounter{figure}
\vfill
\clearpage

\begin{figure}[p]
~\epsfig{file=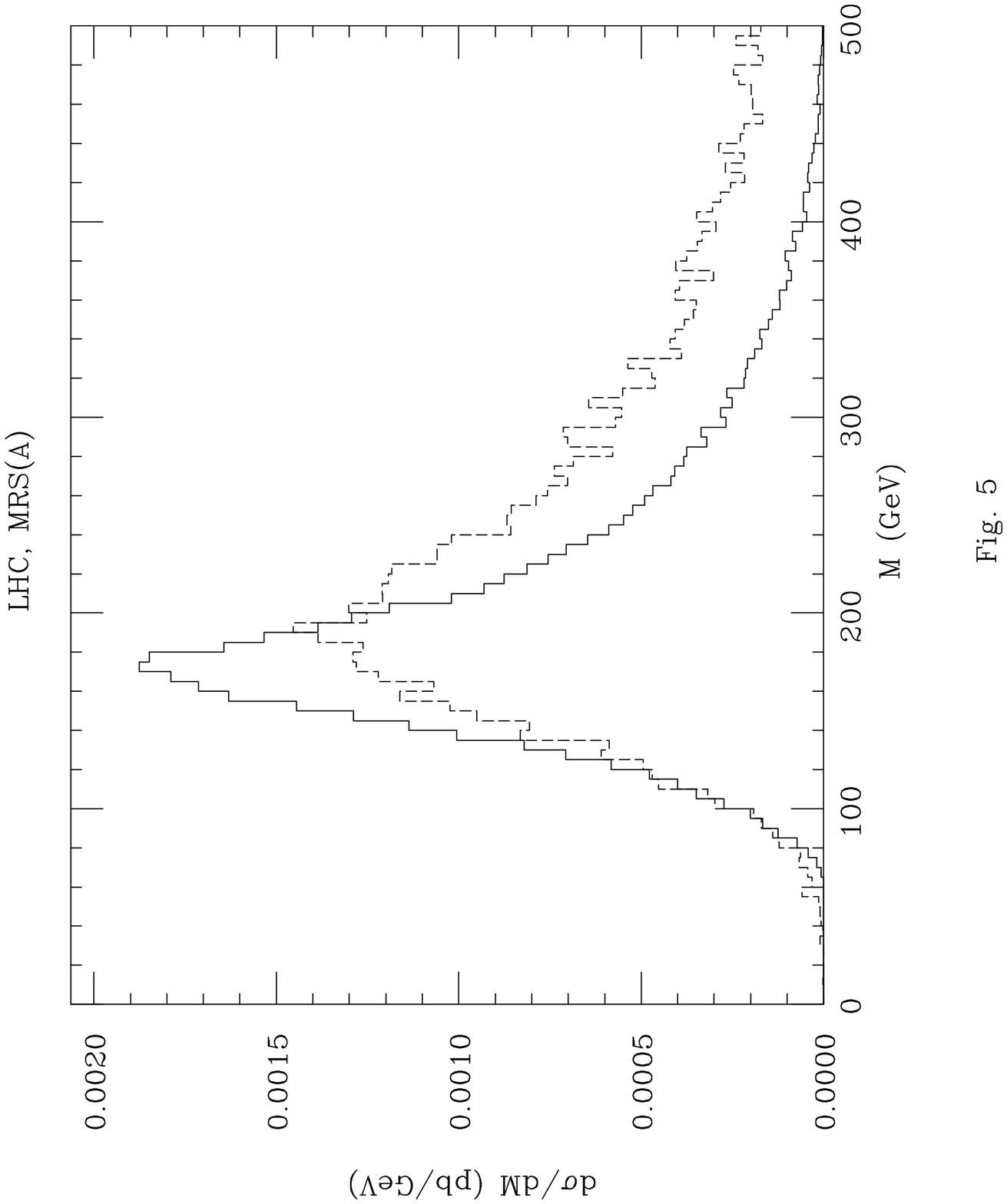,angle=180,height=22cm}
\end{figure}
\stepcounter{figure}
\vfill
\clearpage

\begin{figure}[p]
~\epsfig{file=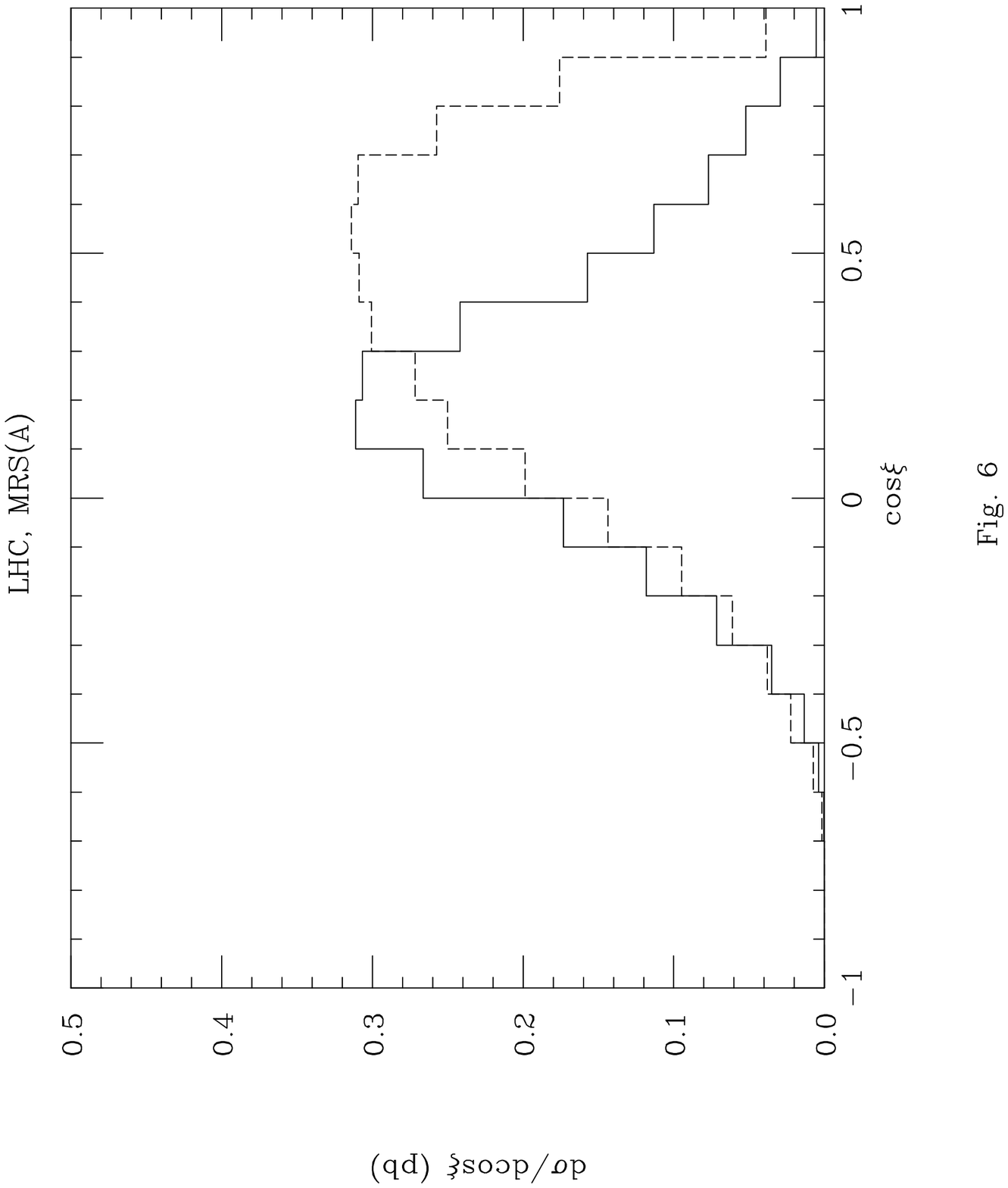,angle=180,height=22cm}
\end{figure}
\stepcounter{figure}
\vfill
\clearpage

\end{document}